\documentclass[twocolumn,aps,superscriptaddress,nofootinbib,groupedaddress]{revtex4-2}

\usepackage{amsmath,amsthm,amsfonts}
\usepackage{graphicx}
\usepackage{physics}
\usepackage{hyperref}
\usepackage{xcolor}
\hypersetup{
	colorlinks   = true, %Colours links instead of ugly boxes
	urlcolor     = blue, %Colour for external hyperlinks
	linkcolor    = blue, %Colour of internal links
	citecolor    = red   %Colour of citations
}

\graphicspath{{figs2/}}

\newcommand{\nc}{\newcommand}

\nc{\calR}{{\cal{R}}}
\nc{\calP}{{\cal{P}}}
\nc{\cN}{ {\cal{N}} }

\nc{\Mpt}{M_{_{\rm Pl}}^2}
\newcommand{\rchi}{\chi}
\newcommand{\rphi}{\phi}
\newcommand{\rvphi}{\varphi}

\newcommand{\epH}{\epsilon_{_H}}

\begin{document}
	
	\title{Multiple Field Ultra Slow Roll Inflation: 
		\\ \rm Primordial Black Holes From Straight Bulk And  Distorted Boundary
	}		
	\author{Sina Hooshangi}
	\email{sina.hooshangi@ipm.ir}
	%	\affiliation{School of Astronomy, Institute for Research in Fundamental Sciences (IPM), Tehran, Iran, P.O. Box 19395-5531}
	
	\author{Alireza Talebian}
	\email{talebian@ipm.ir}
	%	\affiliation{School of Astronomy, Institute for Research in Fundamental Sciences (IPM), Tehran, Iran, P.O. Box 19395-5531}
	
	\author{Mohammad Hossein Namjoo}
	\email{mh.namjoo@ipm.ir}
	%	\affiliation{School of Astronomy, Institute for Research in Fundamental Sciences (IPM), Tehran, Iran, P.O. Box 19395-5531}
	
	\author{Hassan Firouzjahi}
	\email{firouz@ipm.ir}
	\affiliation{School of Astronomy, Institute for Research in Fundamental Sciences (IPM), Tehran, Iran, P.O. Box 19395-5531}
	
	%	\date{\today}
	
	\begin{abstract}
		We study a model of two-field ultra-slow-roll (USR) inflation  bounded by 
		a  curve in the field space. Curvature perturbations and non-Gaussianities can be enhanced both during the USR phase and from the inhomogeneities at the boundary. We employ the full non-linear $\delta N$ formalism to calculate the probability distribution function (PDF) for curvature perturbation non-perturbatively and show that the non-linear effects can significantly enhance the abundance of the primordial black holes (PBHs). For large curvature perturbations, the PDF has a universal exponential tail, but for the intermediate values, the PDF---and, therefore, the abundance of the PBHs---depend sensitively on the geometry of the boundary.  
	\end{abstract}
	
	%\pacs{98.80.Cq}
	\maketitle
	
	\section{Introduction}
	\label{sec:intro}	
	Inflation is the leading paradigm for the early universe cosmology which is well supported by the cosmological observations \cite{Planck:2018jri, WMAP:2010qai}. 
	An almost universal property of the single-field models of inflation 
	is that the amplitude of the local-type non-Gaussianity,  known as the $f_{\rm NL}$ parameter, is at the level of the slow-roll parameters and, therefore, very small. This is known as the non-Gaussianity consistency condition 
	\cite{Maldacena:2002vr, Chen:2010xka}  which provides a relation between the scale dependence of the power spectrum of 
	density perturbations (i.e., the two-point correlation functions) and the amplitude of the three-point functions and $f_{\rm NL}$.  The USR model 
	\cite {Kinney:2005vj}
	is among the very  few known single-field models of inflation which can violate this consistency condition \cite{Namjoo:2012aa, Chen:2013aj, Chen:2013eea, Martin:2012pe}. In the simplest USR setup, the potential is exactly flat, so the inflaton velocity falls off exponentially. As a result, unlike the conventional models, the curvature perturbation on superhorizon scales keeps evolving, leading to the violation of the non-Gaussianity consistency condition.

	The discovery of the gravitational waves from black hole binary mergers by LIGO/VIRGO \cite{LIGOScientific:2016aoc}, inspired attention to the PBHs \cite{Carr:1974nx, Carr:1975qj} as a possible source, which may also contribute to dark matter \cite{Belotsky:2014kca, Sasaki:2016jop, Clesse:2016vqa, Bird:2016dcv, Carr:2016drx, Sasaki:2018dmp, Biagetti:2018pjj, Martin:2019nuw, Fumagalli:2020adf, Carr:2020xqk, Carr:2020gox}. A natural question is under what circumstances an inflationary model can predict a large abundance of PBHs. Again, an attractive---yet simple---possibility is the USR phase of inflation which enhances the typical size of the density perturbations (compared, e.g., to the CMB-scale fluctuations). Besides the enhancement in the power spectrum, it has been noticed that the tail of the PDF of fluctuations can be raised significantly due to the non-perturbative effects in the USR (and, more generally, in the non-attractor) models \cite{Biagetti:2021eep,  Hooshangi:2021ubn, Cai:2021zsp}. This may lead to a drastic change in the PBH formation probabilities. 
	
	Given the significance of the USR model, inspired by the above two applications, it is natural to ask how the predictions are affected if one raises the dimensionality of the field space in a similar setup. In this work, we extend the USR ideas to a two-field scenario with a flat potential. The USR phase  takes place for a few $e$-folds before it ends on a boundary in the field space.  Curvature perturbations can be enhanced not only during the USR phase (i.e., the bulk) but also from the inhomogeneities generated from the boundary of the end of USR. The latter phenomenon was studied in slow-roll inflation in \cite{Lyth:2005qk, Sasaki:2008uc, Naruko:2008sq}. We show that the correlation functions of the curvature perturbation and, more generally, its PDF  depend on the  geometry of the boundary curve as well as the duration of the USR phase, which then lead to non-trivial predictions for the PBHs formation probability.

	\section{The Model and the Background evolution}
	\label{sec:Model}
	
	The setup we consider consists of two scalar fields $\rvphi^a = (\rphi, \rchi)$ minimally coupled to gravity. The universe experiences a phase of slow-roll inflation first---during which the CMB scale perturbations are generated. Then a sudden transition to a short phase of USR occurs, which terminates when the trajectory in the field space hits a boundary. Another phase of slow-roll inflation begins right after, which---we assume---the modes of interest, affected by the USR phase and its boundary, do not evolve significantly until the end of inflation. That is, the adiabatic condition is assumed to be approximately satisfied immediately after the USR phase. A mild transition to the adiabaticity may cause a significant change to the statistics of the curvature perturbation \cite{Cai:2018dkf} which we do not consider in this paper. The location of the USR phase within the whole inflationary trajectory is a freedom in this setup and can be fixed by demanding a specific mass window for the PBHs.
	
	During the USR phase the two fields simultaneously roll on a constant potential and,  assuming an almost constant Hubble expansion rate $H$,
	their background evolutions are  given by the following Klein-Gordon (KG) equation
	\begin{equation} \label{eq:KG}
		\dfrac{\dd^2 \varphi^a}{\dd N^2} + 3 	\dfrac{\dd \varphi^a}{\dd N} \simeq 0\, .
	\end{equation} 
	Here $N$ is the number of $e$-folds, related to the cosmic time $t$  via $\dd N = H \dd t$. Defining the slow-roll parameters via $\epH = -\frac{1}{H}\frac{\dd H}{\dd N}$ and $\eta_{_H} = \frac{1}{\epH} \frac{\dd \epH}{\dd N}$,  the solution of  \eqref{eq:KG} leads to the exponential fall-off of the first slow-roll parameter $\epH \propto e^{-6N}$ (justifying constant $H$ approximation)
	while, accordingly, the second slow-roll parameter is nearly constant, $\eta_{_H} \simeq -6+{\cal O}(\epH) $. Furthermore, since there is no coupling between the fields at the background level, the trajectory of the evolution in the field-space is characterized by a straight line with a slope determined by the ratios of the initial velocity of the background fields, i.e.,
	$\tan \theta = \frac{\dd \bar \chi_i}{\dd N}/\frac{\dd \bar \phi_i}{\dd N}$. Therefore, it is more convenient to rotate the field space so that the new coordinate axes are parallel and normal to the background trajectory (see Fig.~\ref{fig:trajectory}):
	\begin{equation}
		\label{eq:sigma-s}
		\sigma = \cos \theta ~ \phi + \sin \theta ~ \chi \, , \quad s = - \sin \theta ~ \phi + \cos \theta ~ \chi \, .
	\end{equation}
	New fields, $\sigma$ and $s$ are referred as the ``adiabatic" and ``entropy" modes respectively~\cite{Gordon:2000hv}. 
	The solutions of the KG Eqs.~\eqref{eq:KG} then become 
	\begin{equation}
		\label{eq:sigma-N}
		\sigma(N)\simeq \sigma_{\rm i}+\dfrac{\pi_ {\rm i}}{3}\left(1-e^{-3N}\right) \,, \quad
		s = s_{\rm i}  \, ,
	\end{equation}
	where $(\sigma_{\rm i},s_{\rm i})$ are the initial values of the fields while 
	$\pi_{\rm i} = \dd \sigma_{\rm i}/\dd N$ 
	is the initial velocity of $ \sigma$. The above solution indicates that the adiabatic field evolves similarly to the single-field setup \cite{Namjoo:2012aa}. However, nontrivial effects may arise from the surface of the end of USR phase. Unlike the single-field case where  this non-attractor phase ends at a specific point, in the multiple-field scenario it is terminated at a surface determined by the equation $\sigma_{\rm e} = \mathbb{H} (s_{\rm e})$.  As we will see in the subsequent sections, the entropic perturbations contribute to the comoving curvature perturbation ${\cal R}$ only through the surface determined by $\mathbb{H}$. 
	
Note that the boundary $\mathbb{H}$
may come from physical phenomena such as interactions between fields or geometrical features in the field space. An explicit example is the multiple field extension of the hybrid inflation scenario, in which inflation may end when the fields satisfy a certain 
%\blue{\sout{instability}} 
condition 
%\blue{\sout{among the fields}} 
that triggers the instability of the water-fall field, as studied for example in  \cite{Lyth:2005qk, Sasaki:2008uc, Naruko:2008sq}. %\blue{\sout{This feature may also arise when the geometry of the field space has a non-trivial structure. One can imagine that in some part of the field space all fields experience a USR phase during inflation. Then one of the mentioned effects bring inflation to a abrupt slow-roll phase.}}

	The number of $e$-folds from the initial flat hypersurface  to reach the boundary from Eq.~\eqref{eq:sigma-N} is given by
	\begin{equation}
		\label{eq:N-sigma}
		N \big(\sigma_{\rm i}, s_{\rm i} \big)= - \frac{1}{3} \log( 1 + 3 \frac{\sigma_{\rm i} - \sigma_{\rm e}\left(s_{\rm i}\right)}{\pi_{\rm i} }) \,.
	\end{equation}
	This formula resembles the result of the single-field setup in the regime where the classical drift dominates over  the quantum diffusion; with the crucial difference that the additional degree of freedom appears due to the boundary being a curve rather than a point.
	In the drift-dominated regime, the amplitude of stochastic jumps, $ H / 2 \pi$, is small compared to the classical field excursion associated to the classical velocity of the field. Since velocity decays exponentially---to avoid a significant stochastic evolution---it is sufficient to demand that the ratio of the stochastic kicks to the classical velocity is small at the end of USR. This leads to the smallness of the power spectrum of curvature perturbation \cite{Firouzjahi:2018vet} $\sqrt{\calP _{\cal R }}  \ll 1$. 
	Furthermore, the velocity inherited from pre-USR stage during inflation must be the main source of the inflaton  dynamics. For the field to exit the USR phase without the interference  of the  quantum diffusion, one must have $ | \pi_{\rm i}|> 3 |\sigma_{\rm e}-\sigma_{\rm i} | $. These two conditions guarantee that the system does not experience a diffusion-dominated regime during USR phase. In Sec.~\ref{sec:PBH} we shall deal with the rare events that may call into question the validity of the above analysis considering typical realizations. However, it is unlikely that when diffusion is subdominant for the typical events, it contributes to the rare ones, corresponding to the PBH formation,  more than the classical effects. See \cite{Hooshangi:2021ubn} for further discussions. We leave the studies of the diffusion-dominated regime for future work; for relevant works on this direction, see 
	\cite{Pattison:2021oen, Pattison:2017mbe, Ezquiaga:2019ftu, Achucarro:2021pdh}.

	\section{Non-Linear Curvature Perturbation and its Spectra}
	\label{Non_Linear}
	
	To study the statistical properties of the comoving 
	curvature perturbations ${\cal R}$ we use the $\delta N$ formalism \cite{Sasaki:1995aw,Wands:2000dp,Lyth:2004gb,Sugiyama:2012tj,Abolhasani:2019cqw}. 
	To also capture the non-linear effects,  we employ the full non-linear $\delta N$ formalism without Taylor expansion.  Neglecting the fluctuations in the initial velocity of the fields which are diluted rapidly  during expansion, the non-linear ${\cal R}$ may be obtained by perturbing the initial field values, resulting in
	\begin{equation}
		\label{eq:zeta}
		{\cal R}=N \big(\bar \sigma_{\rm i}+\delta \sigma,\bar s_{\rm i} + \delta s \big) - N\big(\bar \sigma_{\rm i},\bar s_{\rm i}\big) \, ,
	\end{equation}
	where an overline denotes the background quantities.
	\begin{figure} 
		\includegraphics[scale=0.75]{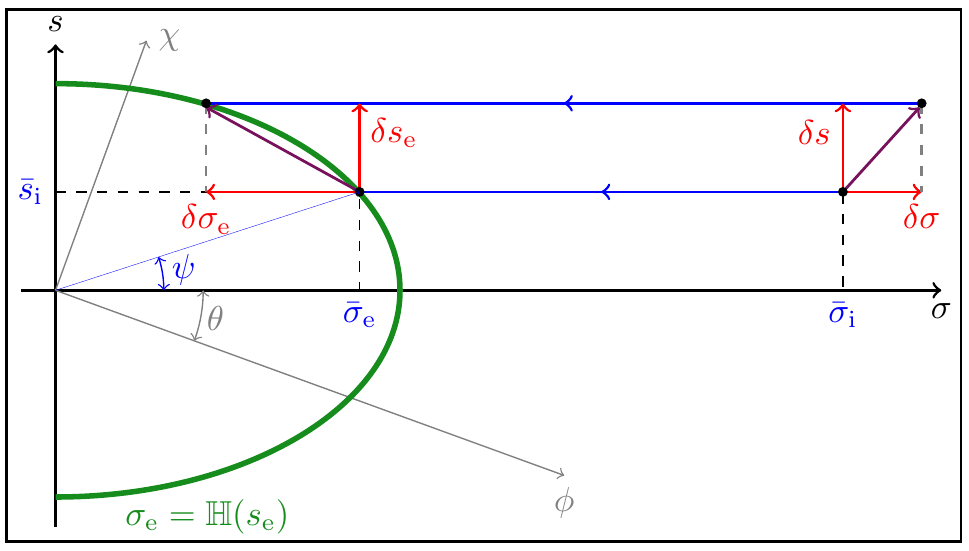}
		\caption{ \footnotesize A schematic view of the USR phase in our setup. Rotating the old coordinates $(\phi, \chi)$ by the angle $\theta$ leads to the adiabatic-entropy coordinates $(\sigma,s)$. The end of USR phase is shown by the thick solid green boundary. The end of unperturbed USR trajectory is parametrized by the angle of intersection $\psi$.}
		\label{fig:trajectory}
	\end{figure}
	As illustrated in Fig.~\ref{fig:trajectory}, the contribution of the entropy fluctuations to ${\cal R}$ is included in the non-linear perturbation of adiabatic fluctuations at the boundary, $\delta \sigma_{\rm e}$. Therefore, $\delta \sigma_{\rm e}$ can be found in terms of the entropy perturbations,
	\begin{equation}
		\label{eq:delta-sigma_e}
		\delta \sigma_{\rm e} = {\mathbb H}(\bar s_{\rm e} + \delta s_{\rm e}) - {\mathbb H}(\bar s_{\rm e}) \equiv h(\delta s_{\rm e}) \,.
	\end{equation}
	Then we immediately find the master equation for ${\cal R}$ as follows:
	\begin{equation}
		\label{eq:zetanl}
		{\cal R}\left(\delta \sigma,\delta s\right) = -\frac{1}{3} \log(1 + 3 \frac{\delta \sigma - h(\delta s)}{\pi_{\rm e}}) \,,
	\end{equation}
	where the single-field USR result can be recovered by $\delta s=0$.
	The non-linear ${\cal R}$ can also be given in terms of the fluctuations in the old coordinates $(\delta \phi,\delta \chi)$, which is somewhat complicated, but it coincides with Eq.~\eqref{eq:zetanl}. 
	
	Having the non-linear relation \eqref{eq:zetanl} at hand one can expand 
	${\cal R}$ to any desired order and calculate its spectra. Our main assumption here is that $(\delta \phi,\delta \chi)$ are uncorrelated and Gaussian field fluctuations with  amplitudes $\Delta \equiv \frac{H}{2 \pi}$.
	This property is passed on to $(\delta\sigma, \delta s)$ due to the linear relation \eqref{eq:sigma-s}.
	
	The dimensionless power spectrum of ${\cal R}$ at the end of USR phase when $N= N_{\rm e}$ is given by
	\begin{align}
		\label{eq:power}
		{\cal P}_{\cal R} ={\cal P}_{\rm SF} \, (1+h'^2) \,;
		\hspace{.5cm}
		h' \equiv \dfrac{{\rm d}h(\delta s)}{{\rm d}\delta s} \bigg|_{\delta s =0} \, ,
	\end{align}
	where ${\cal P}_{\rm SF} =\left(\Delta / \pi_{\rm e}\right)^2$ is the single-field USR counterpart of the power spectrum.
	The above relation indicates that the power spectrum can be enhanced in two different ways.  One is via growing the curvature perturbation during the USR phase controlled by $\pi_{\rm e}$---just like the single-field case---and the other is by the slope of the boundary at the intersection point---which is a genuine feature of the multiple-field scenario. For the USR regime to remain perturbatively under control, $N_{\rm e}$ cannot be arbitrarily large. We may allow for a few $e$-folds of USR phase in the following analysis, corresponding to  
	$N_{\rm e} \lesssim 3$. 
	
	The above two mechanisms of generating curvature perturbations are degenerate at the level of power spectrum. In order to break the degeneracy, we need to investigate the higher spectra. It is straightforward to show~\cite{Lyth:2005fi,Byrnes:2006vq} 
	\begin{align} 
		f_{\rm NL} &= \dfrac{5}{2} + \dfrac{5}{6}\pi_{\rm e}\dfrac{h'^2 h''}{(1+h'^2)^2} \,,
		\\
		\tau_{\rm NL} &= 9 + 6\pi_{\rm e}\dfrac{h'^2 h''}{(1+h'^2)^2} +\pi_{\rm e}^2\dfrac{h'^2 h''^2}{(1+h'^2)^3} \,.
		\label{eq:nong-spectra}
	\end{align}
	Here the primes denote the derivative with respect to the entropy perturbation 
	calculated on the boundary (setting $\delta s=0$ after taking the derivative); $f_{\rm NL}$ measures the amplitude of the three-point correlation 
	function (bispectrum) while $\tau_{\rm NL}$ 
	represents the amplitude of the four-point function (trispectrum)
	\cite{Abolhasani:2019cqw, Byrnes:2006vq} 
	%(see the supplemental material for the other trispectrum parameter $g_{\rm NL}$)
	(see App.~\ref{app:Boundary} for the other trispectrum parameter $g_{\rm NL}$).  The first (constant) terms in  $f_{\rm NL}$ and  $\tau_{\rm NL}$  
	correspond to the bulk (USR) evolution which are the same as in the single-field USR setup, while the remaining terms are the boundary effects. Since the velocity decays rapidly during the USR phase, one may naively conclude that the new terms are sub-dominant. However, as we shall see below,  depending on the properties of the boundary, one can obtain a significant effect from them.
	
	At this step it is worth checking the Suyama-Yamaguchi inequality $\tau_{\rm NL} \geq \left(\frac{6}{5}f_{\rm NL}\right)^2$~\cite{Suyama:2007bg}. For the setup under our consideration we obtain
	\begin{equation}
		\tau_{\rm NL} - \left(\frac{6}{5}f_{\rm NL}\right)^2 = \left(\pi_{\rm e}\dfrac{h' h''}{(1+h'^2)^2} \right)^2 \geq 0 \,.
	\end{equation}
	For the boundaries with $h'=0$ or $h''=0$, the equality is satisfied as in the case of single-field USR setup. 
	
	The above analysis was general, valid for any boundary. We comment that the boundary can take any smooth shape in the two dimensional field space, closed like a circle or an ellipse or open like a line or a hyperbola. As a simple example, we now consider a circle as the boundary, given by the relation $\sigma_{\rm e}^2+s_{\rm e}^2 = R^2$ in the field space.  In this case, from Eq.~\eqref{eq:delta-sigma_e}, $\delta \sigma_{\rm e}$ is related to the entropy fluctuations by
	\begin{equation}
		h(\delta s) =   \sqrt{R^2 - \left(R  \sin\psi +\delta s \right)^2} -R \cos\psi\, .
	\end{equation}
	The angle $\psi$ is defined via $(\tan\psi=\bar s_{\rm e}/\bar \sigma_{\rm e})$ as illustrated in Fig.~\ref{fig:trajectory}; throughout,  we assume $0<\psi<\pi/2$. Therefore, using Eqs.~\eqref{eq:power}-\eqref{eq:nong-spectra}, the spectra for ${\cal R}$ are given by
	\begin{align}
		{\cal P}_\calR &= {\cal P}_{\rm SF} (1+\tan^2\psi)\, ,
		\\
		f_{\rm NL}  &= \dfrac{5}{2} + \dfrac{5}{6\alpha} \sin\psi \tan\psi \,,
		\\
		\label{tau-nl}
		\tau_{\rm NL} &= 9 + \dfrac{6}{\alpha} \sin\psi \tan\psi + \dfrac{1}{\alpha^2}\tan^2\psi  \,,
	\end{align}
	where the parameter  $\alpha$ is related to the radius of  the boundary by $R \equiv \alpha \, \abs{\pi_{\rm e}} $ and we assumed that $\pi_{\rm e}<0$. We see that for some ranges of $\psi$  the power spectrum and the amplitudes of non-Gaussianities may be predominantly generated from the boundary. This is more pronounced when $\psi \sim \frac{\pi}{2}$.

	As mentioned, the non-linearity of the curvature perturbation also alters the shape of the PDF of ${\cal R}$ which in turn affects the PBH formation during the radiation-dominated universe. We deal with this issue in the next section.
	
	%%%%%%%%%%%%%%%%%%%%%%%%%%%
	\section{PBH Formation}
	\label{sec:PBH}
	
	According to the $\delta N$ formalism, the comoving curvature perturbation ${\cal R}$ on a final surface is expressed non-linearly by the perturbations of initial surface $(\delta \sigma , \delta s)$.  Using the nonlinear expression in Eq.~\eqref{eq:zeta}, we thus can calculate the PDF of ${\cal R}$---which we denote by $\bar \rho_{\cal R}$---without appealing to any Taylor expansion 
	via, 
	\begin{equation}
		\bar \rho_{\cal R} = \int_{-\infty}^{+\infty} \delta_D \left( {\cal R} - {\cal R} \left( \delta \sigma , \delta s\right)\right)\, 	\rho_{\delta \sigma, \delta s}\, \dd\delta \sigma\, \dd \delta s\, ,
	\end{equation}
	where, $\delta_D(.)$ is the Dirac delta function and $\rho_{\delta \sigma, \delta s}$ is the joint PDF for the two random fields $\delta \sigma$ and $\delta s$.
	As mentioned earlier, we assume that the two fields $\delta \sigma$ and $\delta s$ are Gaussian and uncorrelated 
	\begin{equation}
		\rho_{\delta \sigma, \delta s} = \frac{1}{2 \pi \Delta^2} \exp(-\frac{\delta \sigma ^2 + \delta s^2}{2 \Delta^2}) \, ,
	\end{equation}
	where, recall,  $\Delta=H/2 \pi$ is the square root of the variance. To compute the probability densities associated to $\cal R$, a subtlety arises due to the fact that not all perturbations in $\delta \sigma$ and $\delta s$ lead to a trajectory along which the fields roll on the USR region and hit the boundary. This may cause an eternal inflation to occur in which case---among other problems---the modes under consideration would not be observable. We thus assume, a priori, that inflation is not eternal so the probability density that we aim to compute is the PDF of $\calR$ {\it conditioned} on the trajectories that cross the boundary. We shall denote that conditional probability density by $\rho_\calR$. Notice that in the eternal inflation regime stochasticity may dominate the dynamics, so excluding the eternal inflation avoids---at least partially---the stochasticity as well. 
	
	%%%%%%%%%%%%%%%%%%%%%%%%%%%
	\begin{figure}
		%	\centering
		\includegraphics[width=.9\columnwidth]{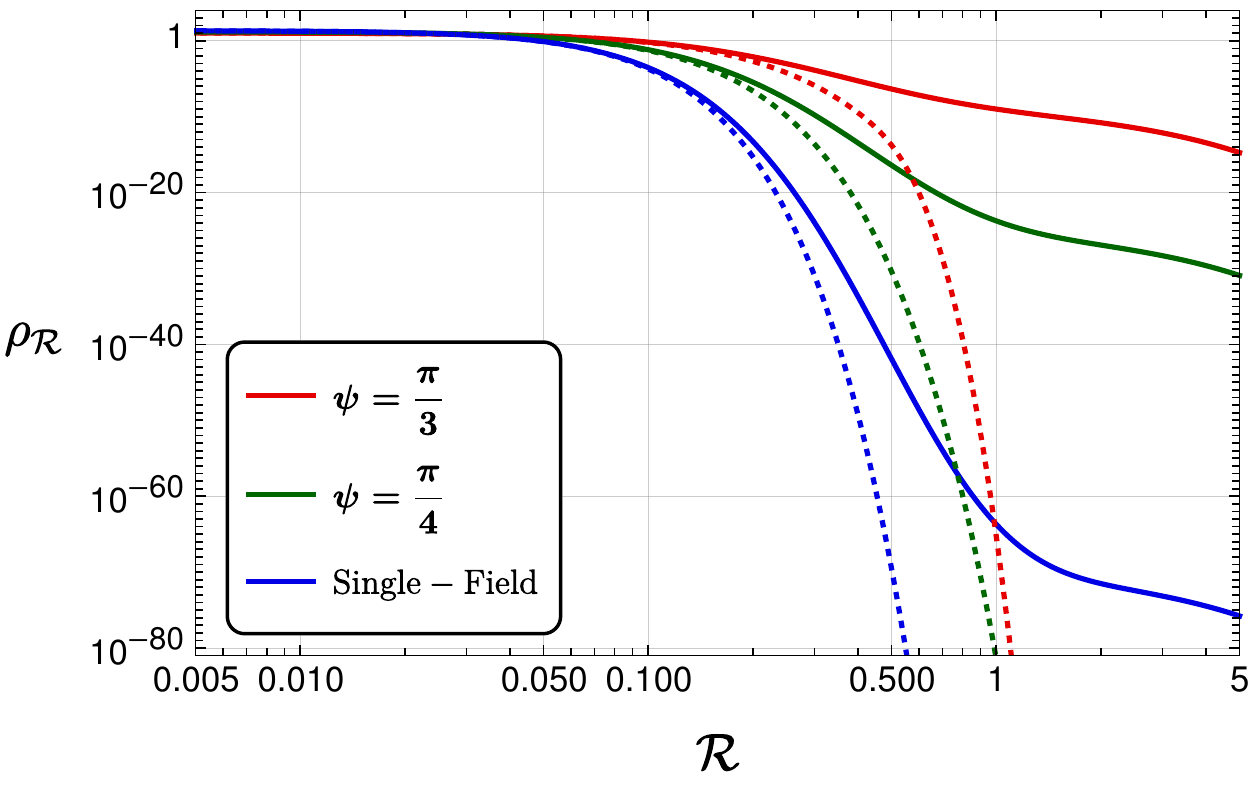}
		\caption{\footnotesize PDF of ${\cal R}$ for a circular boundary calculated from Eq.~\eqref{eq:PDF-zeta-general-bndry} with  $N_{\rm e}=2$ for different values of $\psi$. Dotted lines are obtained by Taylor expanding Eq.~\eqref{eq:zetanl} up to  second order. 	}
		\label{fig:PDF-zeta}
	\end{figure}
	%%%%%%%%%%%%%%%%%%%%%%%%%%%
	
	From the standard relations for the conditional probabilities, we thus obtain  
	\begin{equation} \label{eq:formal-pdf-zeta}
		\rho_\calR = \langle \delta_D \left( {\cal R} - {\cal R} \left( \delta \sigma , \delta s\right)\right) \rangle_{\mathcal{B}}\, ,
	\end{equation} 
	in which, for an arbitrary function $f$, we define 
	 	\begin{equation}
	\langle f	\rangle_{\mathcal{B}}\equiv \dfrac{\int_{\mathcal{B}}f\, \rho_{\delta \sigma, \delta s} \, \dd \delta \sigma \,\, \dd \delta s}{\int_{\mathcal{B}} \, \rho_{\delta \sigma, \delta s} \, \dd \delta \sigma \,\, \dd \delta s }\, ,
\end{equation}
where $\int_\mathcal{B}$ denotes that the integral is taken over the range of $(\delta \sigma, \delta s)$ that the boundary crossing is possible---which shall be discussed in some details shortly. For $f=\delta_D \left( {\cal R} - {\cal R} \left( \delta \sigma , \delta s\right)\right)$---which is the case of interest---we may simplify the numerator by using
\begin{equation}
	\delta_D \left( {\cal R} - {\cal R} \left( \delta \sigma , \delta s\right)\right) = 
	\dfrac{\delta_D \left( \delta \sigma - \delta \sigma \ast \right)}{\left| \partial_{\delta \sigma} \cal R \right|_{\delta \sigma = \delta \sigma \ast}}\, ,
\end{equation}
where, from Eq.~\eqref{eq:zetanl} we have 
\begin{equation}
	\delta \sigma_{\boldsymbol{*}}({\cal R},\delta s) \equiv h(\delta s) + \frac{\pi_{\rm e}}{3} \left(e^{-3 {\cal R}}-1\right)\,  ,
\end{equation}
and 
$\left| \partial_{\delta \sigma} \cal R \right|_{\delta \sigma = \delta \sigma \ast} = \dfrac{e^{3 \cal R}}{\left| \pi_{\rm e}\right|}$.
%	\textcolor{orange}{
%	\begin{equation}
%	\begin{aligned}
%		\rho_{\cal R} &= \dfrac{\int_{\delta s_{\mathrm min}}^{\delta s_{\mathrm max}} \int_{\delta \sigma_{\mathrm min}}^{\delta \sigma_{\mathrm max}} \dd \delta s ~ \dd \delta \sigma ~ \rho_{\delta \sigma , \delta s} \delta_D \left( {\cal R} - {\cal R}(\delta \sigma , \delta s) \right)}{\int_{\delta s_{\mathrm min}}^{\delta s_{\mathrm max}} \int_{\delta \sigma_{\mathrm min}}^{\delta \sigma_{\mathrm max}} \dd \delta s ~ \dd \delta \sigma  \rho_{\delta \sigma , \delta s}} \\
%		&= \dfrac{\int_{\delta s_{\mathrm min}}^{\delta s_{\mathrm max}} \int_{\delta \sigma_{\mathrm min}}^{\delta \sigma_{\mathrm max}} \dd \delta s ~ \dd \delta \sigma ~ \rho_{\delta \sigma} \rho_{ \delta s} \dfrac{\delta_D \left( \delta \sigma - \delta \sigma \ast \right)}{\left| \partial_{\delta \sigma} \cal R \right|_{\delta \sigma = \delta \sigma \ast}} }{\int_{\delta s_{\mathrm min}}^{\delta s_{\mathrm max}} \int_{\delta \sigma_{\mathrm min}}^{\delta \sigma_{\mathrm max}} \dd \delta s ~ \dd \delta \sigma  \rho_{\delta \sigma} \rho_{ \delta s} } \\
%			& = \dfrac{\int_{\delta s_{\mathrm min}}^{\delta s_{\mathrm max}} \dd \delta s ~ \dd \delta \sigma ~ \dfrac{\rho_{\delta \sigma \ast}}{\left| \partial_{\delta \sigma} \cal R \right|_{\delta \sigma = \delta \sigma \ast}} \rho_{ \delta s} }{\int_{\delta s_{\mathrm min}}^{\delta s_{\mathrm max}} \int_{\delta \sigma_{\mathrm min}}^{\delta \sigma_{\mathrm max}} \dd \delta s ~ \dd \delta \sigma  \rho_{\delta \sigma} \rho_{ \delta s} }.	
%	\end{aligned}
%	\end{equation}} 
Putting these  together,  Eq.~\eqref{eq:formal-pdf-zeta} then yields
\begin{align}\label{eq:PDF-zeta-general-bndry}
	\rho_{{\cal R}} = |\pi_{\rm e}| e^{-3 {\cal R}} \frac{\int_{\mathcal{B}_{s}} \exp \Big(-\frac{ \delta s^2 + \delta \sigma_{\boldsymbol{*}}^2 }{2 \Delta^2} \Big) \dd \delta s}{\int_{\mathcal{B}} \exp \Big(- \frac{ \delta s^2 + \delta \sigma^2 }{2 \Delta^2} \Big) \, \dd \delta \sigma \, \dd \delta s} \, ,
\end{align}
where 
$\int_{\mathcal{B}_s}$ indicates that the limits of integral only depend on $\delta s$ (since the integral over $\delta \sigma$ is performed). The integral in the denominator changes the normalization (and significantly deviates from $2\pi \Delta^2$ only if one assumes a background trajectory next to the disallowed regions---which we do not consider).
	
	Note for the large values of ${\cal R}$, the PDF \eqref{eq:PDF-zeta-general-bndry}  behaves as $e^{- 3 {\cal R}}$ independent of the shape of the boundary. At first thought, it seems that the boundary and its geometrical properties may thus not be important for PBH formation. However, we will see that the transition to the above mentioned exponential tail does depend on the geometry of the boundary of the USR phase. Correspondingly, the PBH abundance predicted for various geometries may differ by many orders of magnitudes. %\blue{\sout{ We emphasize again that the exponential tail of large $\cal R$ is merely from the evolution in the USR phase while the transition to the post USR phase is instantaneous.}}
	
	In a two-field setup, besides fixing the boundary, we need to fix four additional  freedoms to fully determine the background evolution. Since  the coordinates are rotated so that the trajectory is along the $\sigma$-axis, we have already set $\frac{\dd s_i}{\dd N}=0$ (see Fig.~\ref{fig:trajectory}). The intersection angle, $\psi$, is the other parameter we use which determines $\bar s_i$ (and recall that $\bar s_i=\bar s_{\rm e}$). Moreover, we  require the prior-to-USR power spectrum to be CMB-compatible, i.e., ${\cal P}_{{\cal R}_i} = 2.1 \times 10^{-9}$. 
	Considering $\pi_{\rm i}^2= 2 \times 10^{-4} \, M_{\rm Pl}^2$ for the initial velocity, we obtain a constant potential with the height $V_0 = 12\pi^2 \Delta^2 \sim 5 \times 10^{-11} M_{\rm Pl}^4$ in which $M_{\rm Pl}$ is the reduced Planck mass.
	Finally, we consider  the total number of $e$-folds during the USR phase, $N_{\rm e}$, as one of our parameters that we vary.  According to the background solutions Eq.~\eqref{eq:sigma-N}, this determines $\bar \sigma_i$ for fixed $\psi$. Therefore, in what follows, we study the predictions of our model, by varying $\psi$ and  $N_{\rm e}$ as two degrees of freedom.
	
	As for the boundary, we mainly consider a circle with the radius $R$ parameterized via $R = \alpha \, \abs{\pi_{\rm e}} $ (notice that $\pi_{\rm e}$ is determined via $\pi_{\rm e}=\pi_{\rm i}e^{-3N_{\rm e}}$).  We only consider the case $\alpha=1$ in this paper. This choice, besides allowing the boundary to play a notable role, also guarantees that the scale of boundary $R$ is larger than the quantum jumps (with the typical size of $\Delta$), so that the stochasticity can be neglected.

	For a fixed value of $\delta s$, the boundary crossing condition for the adiabatic mode is 
	\begin{equation}
		\frac{3}{|\pi_{\rm e}|} (\delta \sigma-h(\delta s)) \in [1-e^{3 N_{\rm e}},1 ] \quad \text{for fixed}\, \delta s
	\end{equation} while for a circular boundary we require  
\begin{equation}
	\delta s+R \sin \psi \in [-R,R] \quad \text{for circular boundary}
\end{equation} for the entropy mode 
%(see the supplemental material  for details)
(see App.~\ref{app:Boundary} for details). These conditions determine the limits of the integrals in Eq.~\eqref{eq:PDF-zeta-general-bndry} which may be performed numerically.
	
	In Fig.~\ref{fig:PDF-zeta} we have plotted $\rho_{{\cal R}}$ for different values of $\psi$. For comparison, we 
	also Taylor expand Eq.~\eqref{eq:zeta} up to the second order and then calculate $\rho_{{\cal R}}$ and also show the single-field non-perturbative results. The results demonstrate the importance of the  non-linear treatment of ${\cal R}$ for PBH formation. This proves the effectiveness of $\delta N$ formalism which captures the full non-linear effects in the classical regime. Compared with the single-field scenario, we also see that the probabilities are enhanced, as a result of the boundary.

Having obtained the PDF for $\calR$, one---in principle---can compute the mass function for PBHs by relating $\calR$ to the density contrast \cite{Musco:2018rwt}.  This is a well-known procedure which can be followed, now that the full PDF of $\calR$ is computed (see App.~\ref{app:beta} where this is outlined for the simplified case of a linear boundary). However---to keep things simple---as a {\it proxy} for the PBH abundance, here we calculate the parameter $\beta$, which is the probability that ${\cal R}>{\cal R}_c$  for some critical value ${\cal R}_c$. Although there are some subtleties regarding ${\cal R}_c$ \cite{Carr:2020gox}, we simply take ${\cal R}_c=1$.
	
\begin{figure}
		\includegraphics[width=.9\columnwidth]{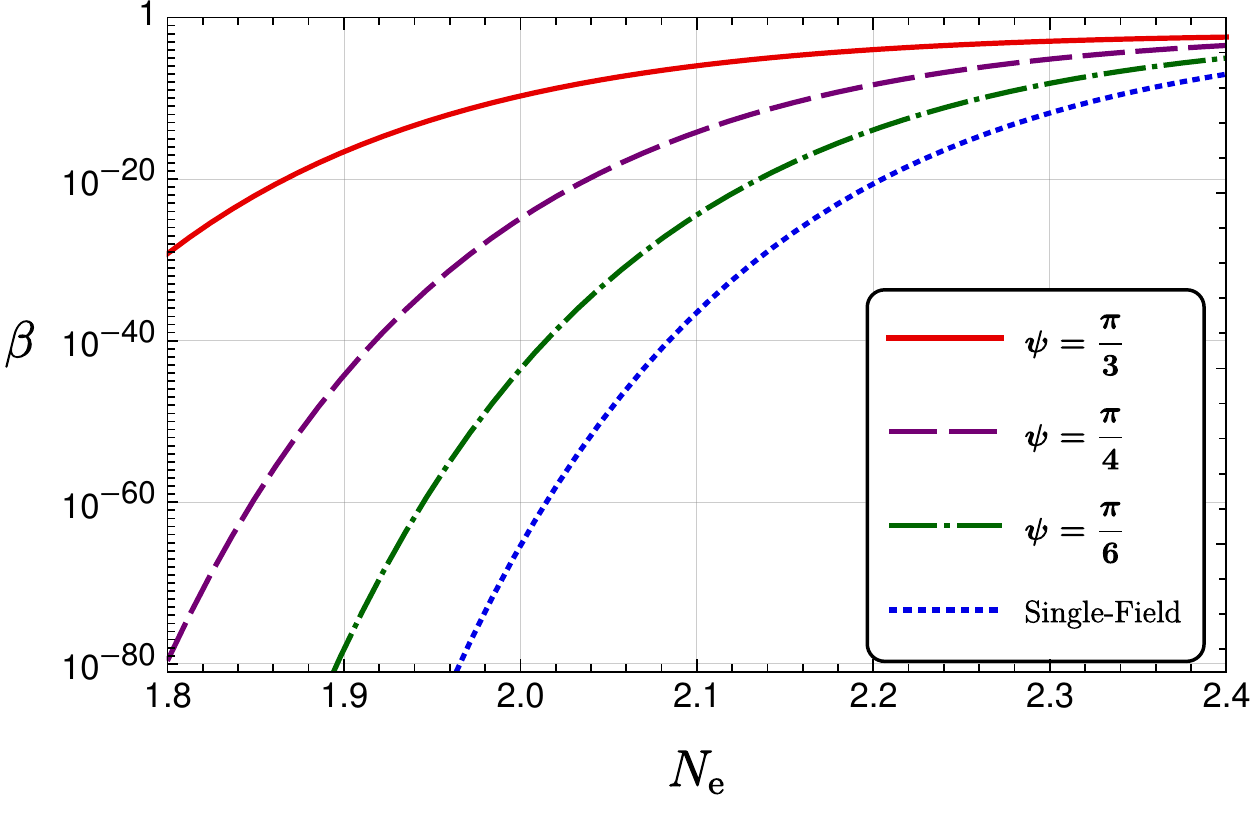}
		\caption{\footnotesize PBH abundance $\beta$ for a circular boundary  for different values of $\psi$ vs. the duration of the USR phase $N_\mathrm{e}$. The differences are more significant for smaller values of $N_\mathrm{e}$}
		\label{fig:beta-Ne-psi}
	\end{figure}

	Fig.~\ref{fig:beta-Ne-psi}  shows   $\beta$  for different angle of intersection $\psi$ when the boundary is a  circle. We also show the results for the single-field case for comparison.

	%%%%%%%%%%%%%%%%%%%%%%%%%%%
	\begin{figure}
		\includegraphics[width=.9\columnwidth]{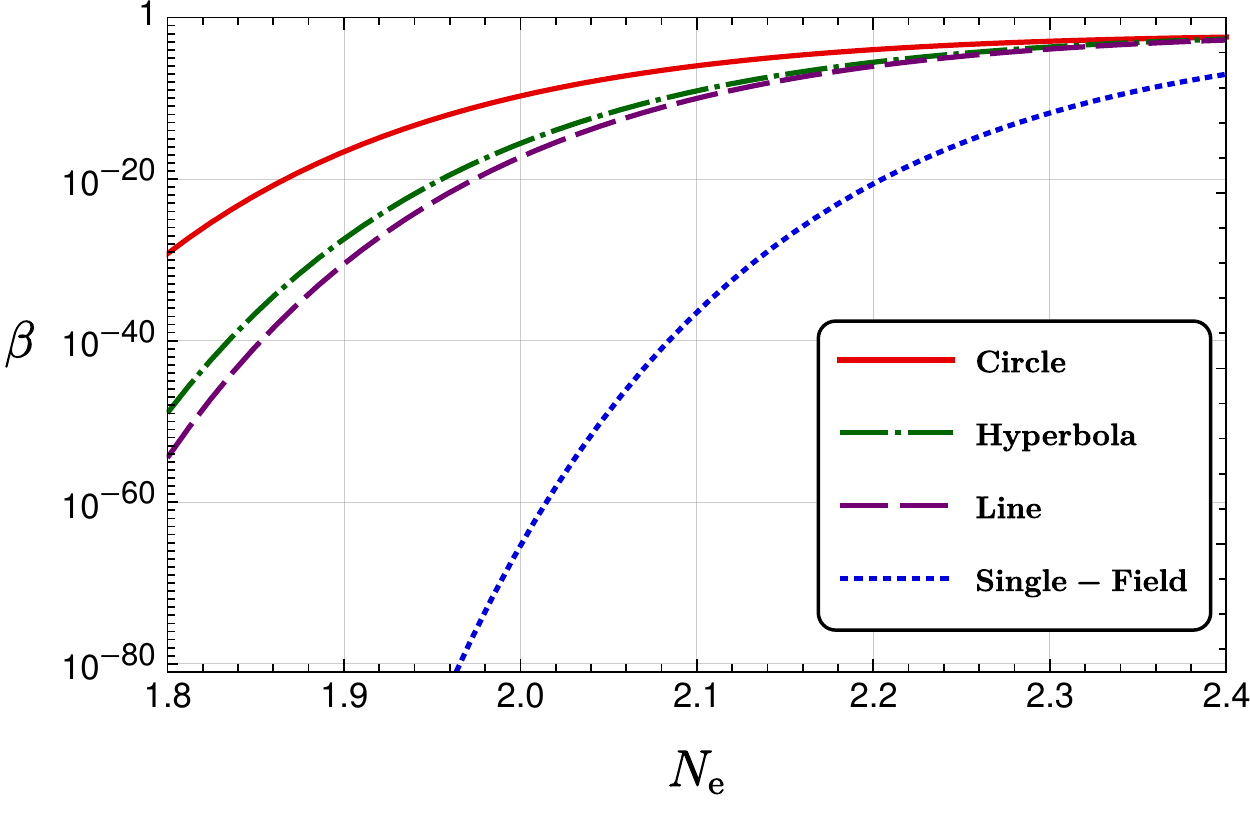}
		\caption{\footnotesize  The parameter $\beta$ when $\psi=\frac{\pi}{3}$ for different boundaries.	}
		\label{fig:beta-boundary}
	\end{figure}
	%%%%%%%%%%%%%%%%%%%%%%%%%%%
	
Furthermore, Fig.~\ref{fig:beta-boundary}  shows how different boundaries affect the PBH abundance. In addition to the circle, we also consider  a hyperbola and a line as the boundaries. We demand that the hyperbola (with the equation $c^2(\sigma_{\rm e}-\sigma_0)^2- s_{\rm e}^2=R_h^2$) and the line are tangent to the circle at the point where the background trajectory hits the boundary (see App.~\ref{app:Boundary} for details). This guarantees that the initial condition parameters and the power spectrum given by \eqref{eq:power} are equal for all boundaries and the differences in the PBH abundance come purely from the geometry of the surfaces. We found that under these assumptions, the PBH abundance is maximal for a circular boundary. This is because the curvature of the circle is larger which implies that larger $\delta N$ is possible for smaller (and more probable) $\delta s$.
	
	\section{Discussions}
	\label{sec:Discussion}
	We have studied a two-field model of USR inflation bounded by various curves in the field space and have shown that  non-trivial effects are generated from the inhomogeneities at the boundary.
	We have employed the $\delta N$ formalism in its full non-linear form to calculate the PDF of the curvature perturbation and also the resulting poly-spectra. We have shown that while the PDF  has a universal tail for large values of ${\cal R}$, i.e., $\rho_{{\cal R}} \propto e^{-3 {\cal R}}$,  for the intermediate values of ${\cal R}$ the PDF---and  hence the abundance of the PBHs---sensitively depend on the geometry of the boundary.
	
	Our analyses can be extended to arbitrary multiple field scenarios with higher dimensional boundaries 
	in the field space to explore how the PDF and statistics of $\cal R$ are sensitive to the dimensionality of the field space. Furthermore, we considered the drift-dominated case, but it would be interesting to study the case where the quantum diffusion becomes important during the USR and on the boundary. We leave addressing these questions  to future works. 
	
	\begin{acknowledgments}
		We thank Amin Nassiri-Rad and Mahdiyar Noorbala for useful discussions.  
	\end{acknowledgments}
	
	\vspace{0.5cm}
	
	\appendix
	
	\section{Boundaries and the boundary crossing conditions}\label{app:Boundary}
	Here, we present some details of how different boundaries---which are tangent to each other---are obtained and how the boundary crossing-criteria put limits on the allowed range of $\delta \sigma$ and $\delta s$. 
	
\begin{figure}[t]
		\begin{center}
			\includegraphics[width=.8\columnwidth]{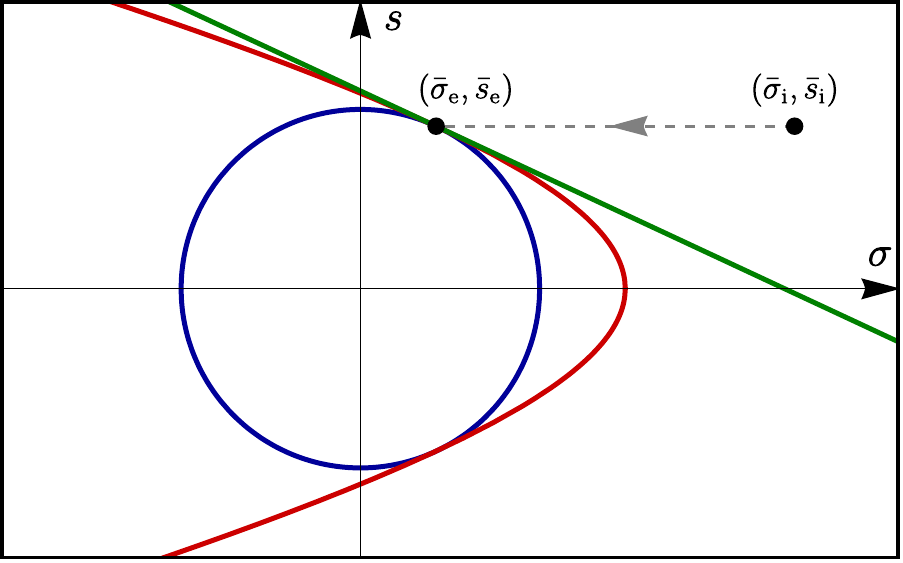}
		\end{center}
		\caption{\footnotesize  A sketch of the classical trajectory hitting different boundaries.
		}
		\label{fig:tangent}
	\end{figure}

	First consider a line with the equation $s_{\rm e}=a \, \sigma_{\rm e} +b $ (which implies that $h(\delta s)=\delta s/a$). We require that the line and the circle are tangent to each other at the point where the background (unperturbed) trajectory hits the boundary (see Fig. \ref{fig:tangent}). This implies 
	\begin{equation}
	\label{eq:line}
		a=-\cot \psi \, , \qquad b= \frac{R }{\sin \psi}.
	\end{equation}
	
	A hyperbolic boundary with the equation $c^2(\sigma_{\rm e}-\sigma_0)^2- s_{\rm e}^2=R_h^2$ will be tangent to the circle at the conjunction point if 
	\begin{align}
		\sigma_0 &= R \cos\psi -\dfrac{\pi_i}{3}\, , \\ 
		\quad c^2 &=\dfrac{3}{|\pi_i|} R \cos \psi \, ,
		\\
		R_h^2 &=  \dfrac{|\pi_i|}{3}R \cos \psi -R^2 \sin ^2 \psi  \, ,
	\end{align}
	where the chosen value of $\sigma_0$ (which determines the location of the hyperbola's center) guarantees that all possible trajectories (with $\pi_i<0$) hit the branch of the hyperbola that is tangent to the circle.  
	Note also that for a hyperbola we have
	\begin{equation}
		c\,	h(\delta s) =   \sqrt{R_h^2 + \left(R_h  \sinh\psi_h +\delta s \right)^2} -R_h \cosh\psi_h\, ,
	\end{equation}
	which may be inserted into Eqs.~\eqref{eq:power}-\eqref{eq:nong-spectra} to obtain different correlation functions. Here, we have defined $\psi_h$ by the relation $R_h \sinh \psi_h= \bar s_e$.

	We are now prepared to obtain the allowed limits of perturbed fields for a given boundary, needed to determine the limits of the integrals in $\rho_{{\cal R}}$ (Eq.~\eqref{eq:PDF-zeta-general-bndry}). Note that we always perform the integral on $\delta \sigma$ first, so in what follows, we first find the allowed range of $\delta \sigma$ for a fixed value of $\delta s$ and then present the remaining limit on $\delta s$ for different boundaries. 
	
	Consider first a fixed value of $\delta s$ and a generic boundary. In order for the $\delta \sigma$ fluctuations not to be so large to bypass the boundary,  the perturbed initial conditions have to satisfy $(\sigma_i \geq \sigma_{\rm e})$ which in terms of the field fluctuations implies
	\begin{equation}
		\bar \sigma_i +\delta \sigma \geq \bar \sigma_{\rm e} + h(\delta s).
	\end{equation}
	On the other hand ---to avoid eternal inflation---for fixed initial field's velocity $\pi_i$, we need  $\sigma_i$ to be such that the initial velocity suffices to reach the boundary. Using Eq.~\eqref{eq:sigma-N} this requirement yields
	\begin{equation}
		\bar \sigma_i +\delta \sigma +\dfrac{\pi_i}{3} \leq \bar \sigma_{\rm e} + h(\delta s).
	\end{equation}
	Using, Eq.~\eqref{eq:sigma-N} again for the unperturbed trajectory, we have 
	\begin{equation}
		\bar \sigma_{\rm e} =\bar \sigma_i +\dfrac{\pi_i}{3} (1-e^{-3N_{\rm e}})\, ,
	\end{equation} 
	and noting that $\pi_{\rm e} = \pi_i e^{-3N_{\rm e}}$ we end up with
	\begin{equation}
		1 - e^{3 N_{\rm e}} \leq \frac{3}{|\pi_{\rm e}|} (\delta \sigma-h(\delta s)) \leq 1.
	\end{equation}
	Note that the lower bound (by demanding the field fluctuations not to bypass the boundary) may seem less justified (because the trajectories that violate that bound would, in principle, be legitimate but unknown to us as we do not specify post-USR phase of inflation). However, since in our setup and choice of parameters the corresponding $|\delta \sigma_{\rm min}|$  is much larger than the width of the Gaussian PDF of the field fluctuations, $\Delta$, this bound---while making the mathematics more rigorous---is practically irrelevant.
	
	For the limits on $\delta s$, we need to specify the boundary. For a circle, we simply obtain 
	$(-R \leq \delta s +R \sin \psi \leq R) $ as can be seen from Fig.~\ref{fig:trajectory}.
	On the other hand, since---unlike a circle---the linear and hyperbolic boundaries are open, they put no limit on $\delta s$ (because for any value of $\delta s$ there are always allowed trajectories for the appropriate range of $\delta \sigma$). 
	
	Before ending the  paper, let us present the results for the other trispectrum parameter $g_{\rm NL}$ and comment on its implications. We have	\cite{Abolhasani:2019cqw, Byrnes:2006vq} 
	\begin{align}
		g_{\rm NL} &= \dfrac{25}{3} + \dfrac{25}{6}\pi_{\rm e}\dfrac{h'^2 h''}{(1+h'^2)^2} + \dfrac{25}{54} \pi_{\rm e}^2\dfrac{h'^3 h'''}{(1+h'^2)^3} \, ,
	\end{align}
	which for the case of circle yields 
	\begin{align}
		g_{\rm NL} &= \dfrac{25}{3} +\dfrac{25}{6\alpha} \sin\psi \tan\psi + \dfrac{25}{18\alpha^2} \sin^2\psi \tan^2\psi \, .
	\end{align}
	Comparing to Eq.~\eqref{tau-nl}, we see that  for typical values of $\psi$, $g_{\rm NL}$ and $\tau_{\rm NL}$ are at the same order.
	It is well-known that large non-Gaussian curvature perturbations  can induce observable  second order gravitational waves (GWs)~\cite{Cai:2018dig,Unal:2018yaa,Adshead:2021hnm}. To be consistent, one has to compare the amplitudes of the trispectrum in various shapes, i.e., the
	$\tau_{\rm NL}$ and  $g_{\rm NL} $ parameters.  Specifically, since the contribution of $g_{\rm NL}$  in the amplitude of the induced GWs can be comparable to that of  $\tau_{\rm NL}$ and $f_{\rm NL}^2$, one may not be justified to simply use the standard perturbative treatment studied before~\cite{Cai:2018dig,Unal:2018yaa}. 
	We leave the investigation of  GWs induced by non-linear curvature perturbations to the future work.

\section{PDF of  the density contrast}\label{app:beta}
In this appendix we outline the derivation of the PDF for the density contrast from which one can compute the PBH mass function. The non-linear relations between the density contrast and the curvature perturbation is discussed e.g., in Ref.~\cite{Musco:2018rwt} where the procedure of deriving the mass function has also been discussed. Here, we will only present the analysis for the simple case where a line represents the boundary of the USR phase, in which case, the analytic computation of the probability distribution of the density contrast would be possible.

We follow the approach outlined and the notation used in Ref.~\cite{Biagetti:2021eep} (and skip the details that can be found there). The key variable for determining the PBH abundance is the following:
\begin{equation}
	\delta_l = - \dfrac{4}{3} r_m \zeta'(r_m)\, ,
\end{equation}
where prime here denotes the derivative with respect to the radial coordinate $r_m$.
In our setup, from Eq.~\eqref{eq:zetanl}, this reduces to:
\begin{equation}
	\delta_l = \dfrac{4}{3} r_m ~ \dfrac{\delta \sigma'/\pi_e - \delta s'/\pi_e ~ \dd h / \dd \delta s }{1 + 3 \delta \sigma /\pi_e - 3 h(\delta s)/\pi_e}\, ,
\end{equation}
where $\delta \sigma' \equiv \delta \sigma'(r_m)$ and so on. Considering a line as the boundary, we have:
\begin{equation}
	\delta_l = \dfrac{4}{3} r_m ~ \dfrac{\delta \sigma'/\pi_e +  \tan\psi \, \delta s'/\pi_e  }{1 + 3 \delta \sigma /\pi_e + 3 \tan\psi \, \delta s/\pi_e}\, ,
\end{equation}
where, similar to App.~\ref{app:Boundary}, we have assumed that the line is tangent to the circle so Eq.~\eqref{eq:line} holds. 
Both numerator and denominator in the above relation are Gaussian random fields since they are summation of Gaussian fields $\delta \sigma'$ and $\delta s'$ or $\delta \sigma$ and $\delta s$. Therefore, we may rewrite our main variable as $\delta_l \equiv \dfrac{X}{Y}$ where $X$ and $Y$ are two uncorrelated, Gaussian random fields with mean $0$ and $1$, respectively, and variances
%
%\begin{figure}[t]%
%	\includegraphics[scale=0.6]{pdf-line.pdf}%
	%\caption{PDF of the density contrast when $\sigma_X^2 \cos^2\psi  = 0.01$.}
%	\label{fig:P_delta_l}
%\end{figure}
%
\begin{align}
	\sigma_X^2 &= \dfrac{1}{\cos^2\psi} \dfrac{16 r_m^2}{9 \pi_e^2} \int \dd \ln k \,\, k^2 \,\, \mathcal{P}(k) \,,%\quad , \quad 
	\\
	\sigma_Y^2 &= \dfrac{1}{\cos^2\psi} \dfrac{9}{\pi_e^2} \int \dd \ln k \,\,  \mathcal{P}(k) \,.
\end{align}
These are exactly relation (19) of Ref.~\cite{Biagetti:2021eep} with the extra factor $\cos^{-2}\psi$. Therefore, the case under study can be obtained simply from the results of \cite{Biagetti:2021eep} by a rescaling of $\pi_e$. That is, when the boundary is linear, the multiple field scenario is degenerate with the single field one. 
And the PDF is simply given by:
\begin{align}
	\rho_{\delta_l} (\delta_l) &=\int \delta_D(\delta_l-X/Y) \rho_X \rho_Y \, \dd X \dd Y \nonumber 
	\\
	&= \dfrac{\sigma_X \sigma_Y e^{-\frac{1}{2 \sigma_Y^2}}}{
		%\pi \sigma_X^2 + \pi \delta_l ^2 \sigma_Y^2
	\pi \, \Sigma^2} 
	%\nonumber \\
	%&
	+ \dfrac{ \sigma_X^2  \erf(\frac{\sigma_X}{\sqrt{2} \sigma_Y \Sigma %\sqrt{\sigma_X^2 + \delta_l^2 \sigma_Y^2}
	})e^{-\frac{\delta_l ^2 }{2 %\left(\sigma_X^2 + \delta_l ^2 \sigma_Y^2 \right)
		\Sigma^2}}}{\sqrt{2 \pi} \Sigma^3 %\left(\sigma_X^2 + \delta_l^2 \sigma_Y^2\right)^{3/2}
	} \,,
\end{align}
which coincides with the findings of Ref.~\cite{Biagetti:2021eep}. Here, we have defined $\Sigma^2 \equiv \sigma_X^2 + \delta_l^2 \sigma_Y^2$ and $\erf$ is the Error function. %In Fig.~\ref{fig:P_delta_l} we have plotted $P_{\delta_l} (\delta_l)$ for different values of $\psi$ when $\sigma_X \cos\psi  = 0.1$.
From this PDF one can obtain the PBH mass function which, due to the aforementioned degeneracy, can be obtained from the results of \cite{Biagetti:2021eep}.

%The mass fraction (which we denote by $	\beta_{\mathrm{PBH}}$, not to be confused with $\beta$ in the paper) can be calculated by:
%\begin{equation}
%	\beta_{\mathrm{PBH}} = \int_{\delta_{l,c}}^{4/3} \dd \delta_l \, \dfrac{M}{M_H} P_{\delta_l} (\delta_l) \,,
%\end{equation}
%where $M_H$ is the horizon mass at the formation time and
%\begin{equation}
%	\dfrac{M}{M_H} = \kappa \left[ \delta_l - \dfrac{3}{8} \delta_l^2 - \delta_c \right]^\gamma \,,
%\end{equation}
%with $\kappa = 3.3$, $\gamma = 0.36$ and $\delta_c = 0.59$. The lower bound of integral is related to:
%\begin{equation}
%	\delta_{l,c} = \dfrac{4}{3} \left( 1 - \sqrt{1 - \dfrac{3}{2} \delta_c}\right) = 0.88 \,.
%\end{equation}
%We also consider only one excited mode which the power spectrum we take is $\calP = \calP_{\delta \phi} k_{\ast} \delta \left(k - k_{\ast}\right)$. The PBH mass fraction is illustrated in Fig.~\ref{fig:beta_PBH} for different values of $\psi$.
%\begin{figure}[h]
%	\includegraphics[scale=0.6]{beta-line.pdf}
%	%\caption{Mass Fraction.}
%	\caption{Mass Fraction.}
%	\label{fig:beta_PBH}
%\end{figure}
	
	\bibliography{MultiUSRBib}

\end{document}